\documentclass[a4paper,12pt]{article}
\usepackage[T2A]{fontenc} 
\usepackage[cp1251]{inputenc} 
\usepackage[russian, english]{babel} 
\usepackage[dvips]{graphicx} 
\makeindex 
\usepackage{epsfig} 
\usepackage[dvips]{texdraw} \input{txdtools}
\usepackage{picinpar} 
\usepackage{amsmath} 
\usepackage{amssymb} 
\usepackage{bm} 
\usepackage{indentfirst} 

\begin{document}

\begin{center}
{\bf \large Difference between radiative transition rates\\
\vspace{2mm} in atoms and antiatoms}
\end{center}

\begin{center}
A.D.\,Dolgov$^{a,c,d}$ \footnote{dolgov@fe.infn.it}, I.B.\,Khriplovich$^{a,b}$
\footnote{khriplovich@inp.nsk.su}, and A.S.\,Rudenko$^{a,b}$
\footnote{a.s.rudenko@inp.nsk.su}
\end{center}
\begin{center}
$^a$ Novosibirsk State University, 630090, Novosibirsk, Russia\\
$^b$Budker Institute of Nuclear Physics, 630090, Novosibirsk, Russia\\
$^c$ ITEP, Moscow, 113218, Russia\\
$^{d}$ Dipartimento di Fisica, Universit\`a di Ferrara, and INFN, Sezione di Ferrara
\\ Ferrara, Italy
\end{center}

\bigskip

\begin{abstract}
We demonstrate that $CP$ violation results in a difference of the partial decay
rates of atoms and antiatoms. The magnitude of this difference is estimated.
\end{abstract}

\vspace{8mm}


The $CPT$ theorem guarantees that the masses of a particle and its
antiparticle are equal. In the same way it guarantees that the
imaginary parts of these masses, i.e. the (inverse) total
life-times of a particle and its antiparticle, are equal as well.

It does not follow however from the $CPT$ theorem that the partial decay rates of a
particle and its antiparticle are the same. In fact, these partial decay rates
should be different due to the $CP$ violation. Certainly, this difference is tiny,
together with the $CP$-odd effects. However, $C$ and $CP$ violation could lead to
the predominance of matter over antimatter in the Universe, for a review see e.g.
ref.~\cite{Dol}. Though the difference between the branching ratios is quite small,
the overall effect amounts to 100\%: the whole Universe is either populated by
matter with almost no antimatter at all, or at least this is true for an
astronomically large domain in our neighborhood.

Nevertheless, a possibility still remains that there is a significant amount of cosmological
antimatter, as is argued e.g.  in ref.~\cite{cosm-anti}. There are several satellite~\cite{satellite}
and balloon~\cite{balloon} missions for search of cosmic antinuclei, in particular for anti-$ {He}^4$,
and a few more detectors are in progress~\cite{planned}.

However, the expected flux of anti-helium is very low, if the antimatter domains are
far from us. The 0.511 MeV line from $e^+e^-$-annihilation or 100 MeV continuum from
$p \bar p$-annihilation into pions may also be quite weak, if matter and antimatter
domains are spatially separated. The ideal source of information about cosmic
antimatter would be atomic spectra if the latter were different for atoms and
antiatoms. However, according to the commonly accepted point of view, it is
impossible to distinguish between atoms and antiatoms having in one's possession
only a flux of radiation from electromagnetic transitions in atoms and antiatoms.
Most probably, according to $CPT$ invariance the positions of the energy levels in
atoms and antiatoms are the same. However, the difference of partial radiative decay
widths in atoms and antiatoms, could differ due to $C$ and $CP$ violation. Below we
estimate the magnitude of this effect and find it to be quite small but non-zero.

A difference between the partial decay rates of atoms and antiatoms may appear if
$C$ and $CP$ are both violated. If $C$ is broken but $CP$ is conserved, the decay
rates into channels with fixed spin values of the participated particles can be
different, but the total decay rates summed over spins must be the same. For their
difference $CP$ must be broken as well. Thus, the $CP$-odd effects considered here
are in fact $C$-odd. In other words, $C$ should be broken and $P$ should be
conserved. Then, in virtue of the $CPT$ theorem, these effects are also $T$-odd and,
as we mentioned above,
$P$-even (TOPE). 

Here we discuss the difference of partial radiative
widths in atoms and antiatoms, due to $CP$ violation. For the
simplicity sake, we confine to the hydrogen and antihydrogen
atoms.


In Refs. \cite{Con,Khr} strict upper limits on the parameters of
the TOPE electron-nucleon interaction were obtained from the
limits on the electron and neutron electric dipole moments. From
these strict limits one can conclude that the effect under
discussion is extremely small. More definite estimates of its
magnitude are presented below.

The TOPE electron-proton interaction Hamiltonian can be
conveniently written as follows \cite{Koz}:
\begin{equation} \label{H}
H_{TOPE}=\frac{1}{m^3_p}\left[k_1 \partial_\nu \left(\bar{\psi} \gamma^5
\sigma^{\mu\nu} \psi \right) \bar{\psi}_p \gamma_\mu \gamma^5 \psi_p + k_2
\bar{\psi} \gamma_\mu \gamma^5 \psi
\partial_\nu \left(\bar{\psi}_p \gamma^5 \sigma^{\mu\nu} \psi_p
\right) \right],
\end{equation}
where $m_p$ is the proton mass; $\psi$ and $\psi_p$ are the wave functions of the
electron and the proton, respectively; $k_1$ and $k_2$ are dimensionless constants;
$\gamma^5=-i\gamma^0\gamma^1\gamma^2\gamma^3$,
$\sigma^{\mu\nu}=\frac{1}{2}\left(\gamma^\mu\gamma^\nu-\gamma^\nu\gamma^\mu\right)$.
(To simplify formulas, we have included the factor $m_e/m_p$ at $k_1$, present in
the definition of $H$ used in Ref. \cite{Koz}, into our definition of $k_1$).

The nonrelativistic limit of Hamiltonian (\ref{H}) is sufficient for our purpose. It
is
\begin{equation} \label{H1}
H_{TOPE}=\frac{1}{2m m^3_p}
\left[k_1(\delta_{ik}\delta_{jl}-\delta_{il}\delta_{jk})+k_2\delta_{ij}\delta_{kl}\right]\sigma_{pj}
\phi^\dagger \sigma_k \left(p^{\prime}+p-2\,e A\right)_{l}\phi\,
\nabla_i\delta(\bm{r}),
\end{equation}
where $\bm{\sigma}_p$ and $\bm{\sigma}$ refer to the proton and electron spins,
respectively; $m_p$ and $m$ are the proton and electron masses; $\bf{p}$ and
$\bf{p}^{\prime}$ are the initial and final momenta of electron; $\bf{A}$ is the
vector potential of radiated photon.

The last term in expression (\ref{H1}) generates the $T$-even,
$P$-odd current density
\begin{equation}\label{cur}
J_l= \frac{e}{m m^3_p}
\left[k_1(\delta_{ik}\delta_{jl}-\delta_{il}\delta_{jk})+k_2\delta_{ij}\delta_{kl}\right]\sigma_{pj}
\phi^\dagger \sigma_k \phi\, \nabla_i\delta(\bm{r}),
\end{equation}
resulting in the contact radiation diagram
\begin{figure}[h]
\center
\includegraphics[scale=1.2]{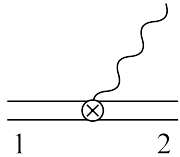}
\caption {Contact $CP$-odd radiation} \label{fig:1}
\end{figure}

\begin{figure}[h]
\center
\includegraphics[scale=1.2]{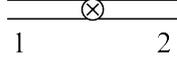}
\caption {$CP$-odd level mixing} \label{fig:2}
\end{figure}

The terms in (\ref{H1}) independent of $\bm{A}$,
\begin{equation} \label{H0}
H_0=\frac{1}{2m m^3_p}
\left[k_1(\delta_{ik}\delta_{jl}-\delta_{il}\delta_{jk})+k_2\delta_{ij}\delta_{kl}\right]\sigma_{pj}
\phi^\dagger \sigma_k \left(p^{\prime}+p\right)_{l}\phi\,\nabla_i\delta(\bm{r}),
\end{equation}
describe the mixing of atomic states 1 and 2 (see
Fig.~\ref{fig:2}). Since interaction $H_0$ is $P$-even scalar, it
mixes only states of the same parity and total angular momentum.

\begin{figure}[h]
\center
\begin{tabular}{c c c}
\includegraphics[scale=1.2]{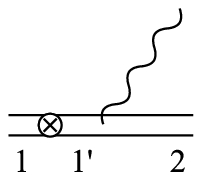} &
\includegraphics[scale=1.2]{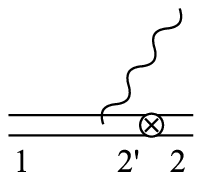} \\
\end{tabular}
\caption {Accompanying $CP$-odd radiation} \label{fig:3}
\end{figure}

\noindent Taken together with the usual electromagnetic interaction, Hamiltonian
(\ref{H0}) generates two more diagrams contributing to the transition amplitude (see
Fig.~\ref{fig:3}).

The discussed $CP$-odd (and $T$-odd) radiation amplitudes are
phase-shifted by $\pi/2$ with respect to the corresponding regular
amplitudes. Therefore, these $T$-odd amplitudes do not interfere
with the regular ones. The corresponding second order contributions
to the decay probabilities are tiny. Moreover, they are the same
for the transitions in atoms and antiatoms.

\begin{figure}[h]
\center
\begin{tabular}{c c c c}
\includegraphics[scale=1.2]{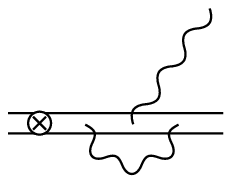} &
\includegraphics[scale=1.2]{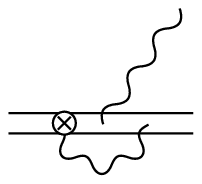} &
\includegraphics[scale=1.2]{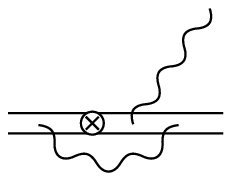} &
\includegraphics[scale=1.2]{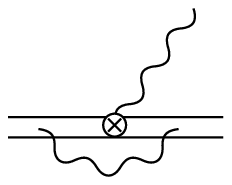}
\end{tabular}
\caption {Loop contributions to the admixed $CP$-odd radiation amplitude}
\label{fig:4}
\end{figure}

\begin{figure}[h]
\center
\begin{tabular}{c c c}
\includegraphics[scale=1.2]{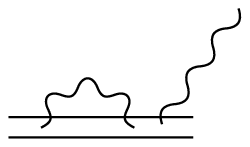} &
\includegraphics[scale=1.2]{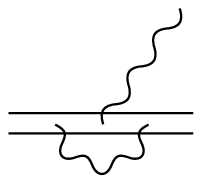} &
\includegraphics[scale=1.2]{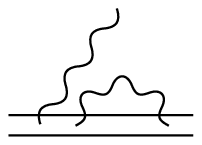} \\
(a) & (b) & (c)
\end{tabular}
\caption {Loop contributions to the regular $CP$-even radiation amplitude}
\label{fig:5}
\end{figure}

However, the difference between the partial decay rates in atoms and antiatoms does
exist. It arises on the loop level due to the imaginary parts of the $CP$-odd and
$CP$-even diagrams presented in Figures~\ref{fig:4} and \ref{fig:5}, respectively.
In Fig.~\ref{fig:4} only some typical diagrams are presented; the total number of
such diagrams is 21, they can be obtained from those in Fig.~\ref{fig:4} by all
possible permutations of vertices.


Simple dimensional estimate for the relative difference of the partial transition
widths $w$'s of atom and antiatom, using Hamiltonian (\ref{H}) and taking into
account the radiative correction presented in Figs. 4 and 5, looks as follows:

\begin{equation}\label{est}
\frac{\Delta w}{w} \sim \left(\frac{m \alpha}{m_p}\right)^3\alpha \,k_{1,2} \sim
10^{-19}\,k_{1,2};
\end{equation}
here $1/m_p^3$ enters Hamiltonian (\ref{H}) explicitly; one more $\alpha$ originates
from imaginary parts of the loop diagrams 4 and 5.

This estimate is quite obvious for the contributions generated by the contact
radiation diagram Fig. 1. The situation with the contributions originating from
the diagrams 3a,b is more subtle. First of all, for the coinciding states, 1 and $1^\prime$
(or 2 and $2^\prime$) the corresponding matrix elements vanish identically. Then, if the primed
and unprimed states, 1 and $1^\prime$ (or 2 and $2^\prime$) are separated by the fine-structure interval
only, one might expect that the effect would be enhanced $\sim 1/\alpha^2$.
In this case, however, the matrix elements of the transitions between the primed
and unprimed states are suppressed $\sim \alpha^2$. Thus, we arrive again at
the same estimate (5).

The present TOPE constants $k_{1,2}$ are related to those used in
\cite{Con, Khr} (see formulas (11.14) and (11.24) in \cite{Khr})
as follows:
\begin{equation}
k_{1,2}=(Gm^2_p/2\sqrt{2})q_{eq,qe}.
\end{equation}
In Ref. \cite{Khr} limits on the parameters of the TOPE
interaction $q_{qe}<10^{-4}$ and $q_{eq}<10^{-7}$ were obtained
(see formula (11.27) therein), which result in $k_2<10^{-9}$ and
$k_1<10^{-12}$. Thus, even under the more liberal assumption
$k_2<10^{-9}$, we arrive at quite impressive upper limit on
the relative difference of the partial transition widths $w$'s in
hydrogen and antihydrogen:
\begin{equation}
\frac{\Delta w}{w} \lesssim 10^{-28}.
\end{equation}

Anyway, the effect exists and though at present it is far
from possible observation, its study deserves attention.

\subsection*{Acknowledgements}
The work was supported in part by the Ministry of Education and Science of the
Russian Federation, by the Foundation for Basic Research through Grant No.
11-02-00792-a, and by the Grant of the Government of Russian Federation, No.
11.G34.31.0047.

\renewcommand{\bibname}

\end{document}